\documentclass{article}
\usepackage{algorithm}
\usepackage{algorithmic}
\usepackage{graphicx}    
\usepackage{subcaption}  
\usepackage[english]{babel}
\usepackage[
  a4paper,
  top=2cm,
  bottom=2cm,
  left=3cm,
  right=3cm,
  marginparwidth=1.75cm
]{geometry}

\usepackage{indentfirst} 
\usepackage{ragged2e} 
\usepackage{amsfonts} 

\justifying 
\usepackage{amsmath} 
\usepackage{hyperref} 
\usepackage{natbib} 

\title{RMD: Robust Modal Decomposition with Constrained Bandwidth}

\author{
  Wang Hao$^{*}$ \quad
  \href{mailto:25b905039@stu.hit.edu.cn}{25b905039@stu.hit.edu.cn} \\
  Kuang Zhang \quad \href{mailto:zhangkuang@hit.edu.cn}{zhangkuang@hit.edu.cn} \\
  Hou Chengyu$^{\dagger}$ \quad \href{mailto:houcy@hit.edu.cn}{houcy@hit.edu.cn} \\
  Yang Yifan \quad \href{mailto:mgmgmgmagic@163.com}{mgmgmgmagic@163.com} \\
  Tan Chenxing \quad \href{mailto:3424611356@qq.com}{3424611356@qq.com} \\
  Fu Weifeng \quad \href{mailto:2298000456@qq.com}{2298000456@qq.com} \\
  \small School of Electronics and Information Engineering, \\
  \small Harbin Institute of Technology, Harbin 150001, China
}

\begin{document}
\maketitle 
\thanks{$^{*}$ First author. $^{\dagger}$ Corresponding author: houcy@hit.edu.cn.}

  \maketitle

  \begin{abstract}
    Modal decomposition techniques, such as Empirical Mode Decomposition (EMD), Variational
    Mode Decomposition (VMD), and Singular Spectrum Analysis (SSA), have advanced
    time-frequency signal analysis since the early 21st century. These methods
    are generally classified into two categories: numerical optimization-based methods
    (EMD, VMD) and spectral decomposition methods (SSA) that consider the
    physical meaning of signals. The former can produce spurious modes due to the
    lack of physical constraints, while the latter is more sensitive to noise
    and struggles with nonlinear signals. Despite continuous improvements in these
    methods, a modal decomposition approach that effectively combines the
    strengths of both categories remains elusive.Thus, This paper proposes a Robust
    Modal Decomposition (RMD) method with constrained bandwidth, which preserves
    the intrinsic structure of the signal by mapping the time series into its
    trajectory-GRAM matrix in phase space. Moreover, the method incorporates
    bandwidth constraints during the decomposition process, enhancing noise resistance.Extensive experiments were conducted to validate its performance: on the synthetic dataset front, we focused on low-SNR sine wave separation tasks and nonlinear signal processing experiments to verify the method’s ability to extract weak signals and handle nonlinear distortions; on real world data, we tested it on a real-collected millimeter-wave micro-motion energy dataset to demonstrate its practical applicability in radar-related micro-motion signature analysis.All code and dataset samples are publicly available on GitHub for reproducibility: https://github.com/Einstein-sworder/RMD.
  \end{abstract}

  \noindent

  \textbf{Keywords:} Modal Decomposition, Eigenvalue Decomposition, Constrained
  Bandwidth,

  Robust Signal Processing

  \section{INTRODUCTION}

  Modal decomposition techniques, represented by methods such as Empirical Mode Decomposition
  (EMD)\cite{1},Variational Mode Decomposition (VMD)\cite{2}, and
  Synchrosqueezed Transform (SST)\cite{3}, are regarded as significant
  advancements in signal time-frequency analysis since the beginning of this
  century. As a generalized adaptive filter\cite{4,5,6}, these methods have shown
  wide-ranging applications in various fields, including physiological signal
  analysis\cite{7,8,9}, meteorological analysis\cite{10,11,12},radar signal processing\cite{13,14,15},
  industrial monitoring systems\cite{16,17},etc.

  Currently, modal decomposition methods are mainly categorized into two classes:
  convex optimization-based decomposition and singular spectrum decomposition. The
  Empirical Mode Decomposition (EMD) method proposed by Huang et al\cite{1} is
  considered a pioneering work in modal decomposition. Following this, various
  methods have been developed, including Ensemble EMD (EEMD), Complete EEMD (CEMD)\cite{18},
  Variational Mode Decomposition (VMD) \cite{2}, Singular Spectrum Analysis(SSA)
  \cite{19,20,21}, Symplectic Geometry Mode Decomposition (SGMD) \cite{22}, and
  Jump Plus AM-FM Mode Decomposition (JOT) \cite{23}. From a mathematical perspective,
  these algorithms can be primarily divided into iterative solution-based
  methods and matrix decomposition-based methods. The iterative methods, such as
  EMD, EEMD, CEMD, and VMD, often use iterative techniques to decompose signals
  into individual modes. For example, the EMD method decomposes a signal through
  successive envelope extraction to obtain different modes, while VMD, based on
  variational optimization, equivalently constructs an adaptive Wiener filter bank
  to separate modes. These methods offer rich adjustable parameters, strong noise
  resistance, and are mathematically intuitive, but they suffer from issues such
  as parameter sensitivity and the potential for spurious components.

  On the other hand, methods based on matrix decomposition, such as SSA and SGMD,
  focus on constructing a trajectory matrix from the signal and mapping it into
  the phase space. Singular value decomposition (SVD) is then used to separate the
  modal components. SGMD performs a symplectic geometric transformation on the
  trajectory matrix before conducting eigenvalue decomposition, maintaining the system's
  symplectic geometry invariance. These methods are generally considered
  effective at preserving the true structure of the signal—especially when the
  signal-to-noise ratio (SNR) is high. However, their performance degrades
  significantly when the SNR is low or when non-stationary interference signals
  are present.

  \begin{figure}[htbp]
    \centering
    \includegraphics[width=0.8\linewidth]{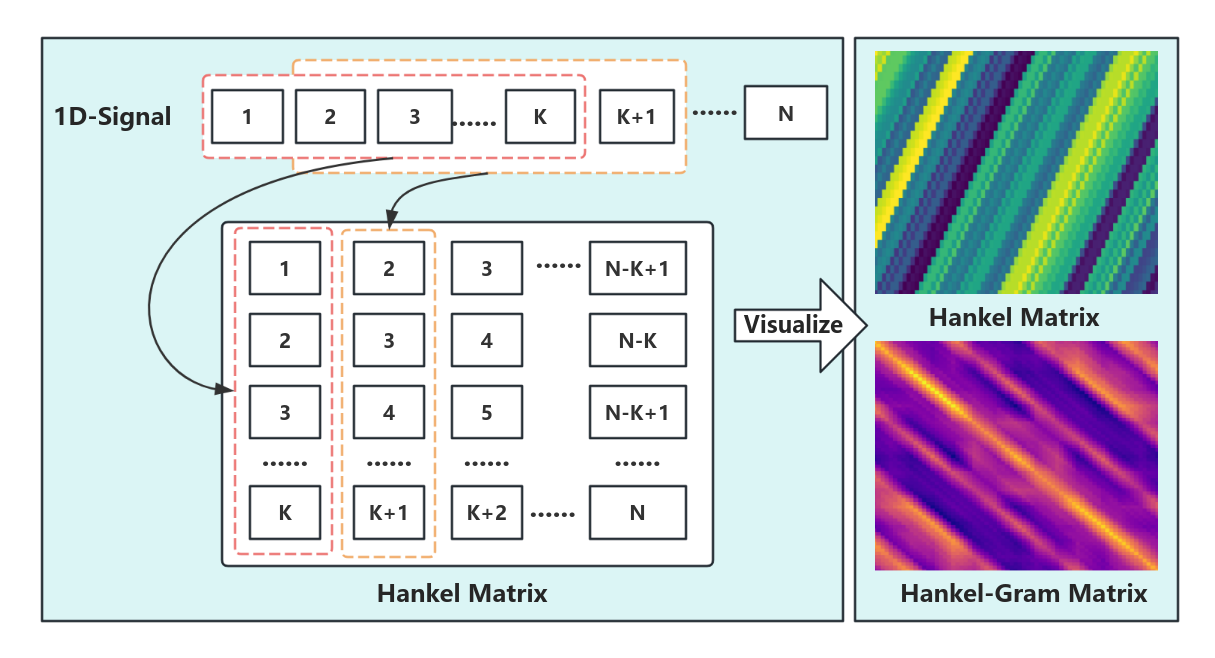}
    \caption{\small It illustrates the process of constructing a Hankel matrix
    with embedding dimension K and delay $\tau$=1 from a one-dimensional signal.
    It also visualizes the Hankel matrix and its Hankel-Gram matrix as heatmaps
    for a data waveform presented later in the paper. From the heatmap, it is
    evident that the Hankel-Gram matrix is smoother and more symmetric compared
    to the Hankel matrix itself \cite{19}.}
    \label{fig:hanckle_gram}
  \end{figure}

  To address the limitations of the aforementioned methods, this paper proposes
  improved method that combines the strengths of both approaches, referred to
  the RMD method. The RMD method performs eigenvalue decomposition of the
  Gram matrix of the signal's trajectory matrix and reconstructs the signal
  using the corresponding eigenvectors. The novel aspect of this approach lies in
  the inspiration drawn from VMD. When solving for the eigenvalues of the Gram
  matrix, first- or second-order difference operators are introduced. By defining
  a unique augmented matrix structure, the method achieves enhanced noise
  suppression, modal bandwidth limitation, and interference mitigation
  capabilities.

  The rest of the paper is organized as follows: Section II reviews the basic
  sign al model, mathematical principles, and fundamental concepts. In Section III,
  we present the underlying idea and detailed steps of the proposed method. Section IV
  provides experimental results on synthetic and real-world data to demonstrate
  the advantages of our approach. Section V offers a conclusion.

  \section{MATHEMATICAL PRINCIPLES}

\subsection{Bandwidth Constraints of Modal Decomposition}

In contrast to the original definition in the EMD paper\cite{1}, most of the works after in the literature define the modal functions in the following form:
\begin{equation}
u_{k}(t) = A_{k}(t)\cos(\phi_{k}(t))
\end{equation}
where $u_{k}(t)$ represents the $k$-th modal signal, with $A_{k}(t)$ being the
amplitude (modulated by $\phi_{k}(t)$), and $\phi_{k}(t)$ representing the phase
modulation of the oscillatory signal.

Taking the derivative of $u_{k}(t)$:
\begin{equation}
u_{k}^{\prime}(t) = A_{k}^{\prime}(t)\cos(\phi_{k}(t)) - A_{k}(t)\phi_{k}^{\prime}(t)\sin(\phi_{k}(t))
\end{equation}
The first term can be interpreted as the amplitude variation, while the second
term represents the frequency modulation.

The instantaneous frequency is given by:
\begin{equation}
\omega_{k}(t) = \phi_{k}^{\prime}(t)
\end{equation}
Assuming that the amplitude variation is slow, i.e., $A_{k}^{\prime}(t) \approx
0$, we have:
\begin{equation}
u_{k}^{\prime}(t) \approx -A_{k}(t)\omega_{k}(t)\sin(\phi_{k}(t))
\end{equation}
Thus, the energy (or $L_{2}$-norm) of the signal is given by:
\begin{equation}
\int |u'_{k}(t)|^{2} dt \approx \int A_{k}^{2}(t) \omega_{k}^{2}(t) \sin^{2}(\phi_{k}(t)) dt
\end{equation}

It follows that if $\omega_{k}(t)$ exhibits small fluctuations, the differential
energy will also be small, corresponding to a narrow bandwidth. In general,
the decomposed modes will not be concentrated around the DC (direct current)
frequency. Therefore, by constraining $\int |u'_{k}(t)|^{2} dt$, modal components with a narrower bandwidth are obtained, as discussed in \cite{2}.

For discrete signals $x[n]$, the constraint should be expressed as:
\begin{equation}
 \nabla u_{k} ||_{2}^{2} = \sum_{n} |u_{k}[n+1] - u_{k}[n]|^{2}
 \label{eq:6} 
\end{equation}

\subsection{PCA Reconstruction in Signal Phase Space}
\subsubsection{Trajectory Matrix Representation of the Signal}

Given a discrete signal $x[n]$, the trajectory matrix (constructed following
the Hankel matrix rule) \cite{19}is expressed as:
\begin{equation}
X =
\begin{bmatrix}
x[1]   & x[2]   & \dots  & x[N-K+1] \\
x[2]   & x[3]   & \dots  & x[N-K+2] \\
\vdots & \vdots & \ddots & \vdots   \\
x[K]   & x[K+1] & \dots  & x[N]
\end{bmatrix}
\end{equation}
where $K$ is the embedding dimension, satisfying $K \leq N$, and $N$ is the
signal length. In this work, we set the delay parameter $\tau = 1$.

According to the time-delay embedding theorem (Takens' theorem), if the system's
intrinsic dimension $d$ satisfies $d \leq K$, the trajectory matrix retains the
topological structure of the original dynamical system. The trajectory matrix
describes the evolution of the signal in the $K$-dimensional phase space.

\subsubsection{Eigenvalue Decomposition of the Trajectory Gram Matrix}

Since the matrix $X$ is generally not square, we perform Singular Value
Decomposition (SVD) on $X$:
\begin{equation}
X = U \Sigma V^{T}
\end{equation}
where $\mathbf{U} \in \mathbb{R}^{L \times r}$ and $\mathbf{V} \in \mathbb{R}^{K \times r}$ are the left and right singular vectors, respectively, and $\mathbf{\Sigma} = \mathrm{diag}(\sigma_1, \sigma_2, ..., \sigma_r)\mathbf{X}_i = \sigma_i \mathbf{u}_i \mathbf{v}_i^T$
is the diagonal matrix of singular values.

To reconstruct the modal components, we use the diagonal averaging method on
the singular vectors:
\begin{equation}
\mathbf{X}_{i} = \sigma_{i} \mathbf{u}_{i} \mathbf{v}_{i}^{T}
\end{equation}
This method is referred to as Singular Spectrum Analysis (SSA) of the signal, but it is sensitive to local disturbances and exhibits poor noise resistance\cite{22}. To overcome these issues, we introduce the Principal Component Analysis (PCA) method, which performs eigenvalue decomposition of the Gram matrix of the trajectory matrix $X$, incorporating the
global autocorrelations of the signal:
\begin{equation}
G = V\Sigma V^T
\end{equation}

Here, $\mathbf{G} = \mathbf{X}^T \mathbf{X} \in \mathbb{R}^{K \times K}$ is the Gram matrix of the trajectory matrix $\mathbf{V} = [\mathbf{v}_1, \mathbf{v}_2, ..., \mathbf{v}_K]$, and $\phi_{k}$ are
the eigenvectors representing the time evolution modes of the signal in phase
space. Each mode corresponds to a specific dynamic component. The eigenvalue
diagonal matrix $\Lambda = \mathrm{diag}(\lambda_1, \lambda_2, ..., \lambda_K)$ reflects the energy of each mode since, due to the semidefiniteness
of $G$, all eigenvalues $\lambda_{i}$ are non-negative.

By selecting the eigenvectors $\phi_{k}$ corresponding to the largest
eigenvalues, the trajectory matrix can be reconstructed as:
\begin{equation}
Z_i= X\mathbf{v}_i\mathbf{v}^T_i
\end{equation}
Then, the one-dimensional signal components are obtained by diagonal averaging:
\begin{equation}
\mathbf{Z}_{i} = \mathbf{X} \mathbf{v}_{i} \mathbf{v}_{i}^{T}
\hat{s}_{k,p} = \begin{cases} \frac{1}{k} \sum_{m=1}^{k} \hat{x}_{m,k-m+1,p}, & \mathrm{for} 1 \leq k < L^{*} \\ \frac{1}{L^{*}} \sum_{m=1}^{L^{*}} \hat{x}_{m,k-m+1,p}, & \mathrm{for} L^{*} \leq k < K^{*} - 1 \\ \frac{1}{N - k + 1} \sum_{m=k-K^{*}+1}^{L^{*}} \hat{x}_{m,k-m+1,p}, & \mathrm{for} K^{*} \leq k \leq N. \end{cases}
\end{equation}

\subsubsection{PCA Reconstruction Preserves Symplectic Invariance of the Phase Space}

The method of symplectic geometric modal decomposition (SGMD), as proposed in \cite{22},
preserves the symplectic structure of the signal in phase space. We now demonstrate
that the method described in section (2) is fully equivalent to this approach.

Define the symplectic matrix as:
\begin{equation}
\mathbf{J}=
\begin{bmatrix}
0  & I \\
-I & 0
\end{bmatrix}
\end{equation}
where $\mathbf{J}$ is the standard symplectic matrix, satisfying $\mathbf{J}^{T}
= -\mathbf{J}$.

From the trajectory Gram matrix, we construct the Hamiltonian matrix:
\begin{equation}
H =
\begin{bmatrix}
G & 0      \\
0 & -G^{T}
\end{bmatrix}
\end{equation}

SGMD requires finding an orthogonal, anti-symmetric symplectic matrix $Q$,
which transforms $H^{2}$ into a block upper triangular form, thus the
symplectic structure is preserved. The components of $Q$ are then used to reconstruct the
modes:
\begin{equation}
Q^{T}H^{2}Q =
\begin{bmatrix}
B & R     \\
0 & B^{T}
\end{bmatrix}
\end{equation}

Let $\mathbf{X}= U \Sigma V^{T}$ (SVD decomposition), and define:
\begin{equation}
Q = \frac{1}{\sqrt{2}}
\begin{bmatrix}
U & U  \\
V & -V
\end{bmatrix}
\end{equation}

Then, we have:
\begin{equation}
Q^{T}G Q =
\begin{bmatrix}
\Sigma^{2} & 0          \\
0          & \Sigma^{2}
\end{bmatrix}
\end{equation}

Thus, the eigenvectors $U$ obtained via PCA directly generate the orthogonal symplectic
matrix $Q$, which preserves the symplectic invariance of the phase space
structure.

\subsection{Augmented Matrix and Regularization Objective Function}

To limit the bandwidth of modes, inspired by \eqref{eq:6}, we transform the one-dimensional mode bandwidth constraint into an expression in phase space, as stated in Theorem 1:

\subsection*{Theorem 1}

Let $\mathbf{v}_i$ be the eigenvector of $\mathbf{G}$, and the trajectory matrix reconstructing the $i$-th mode is $\mathbf{X}_i = \mathbf{U}_i \mathbf{v}_i^T$, where $\mathbf{u}_i = \frac{\mathbf{X} \mathbf{v}_i}{\|\mathbf{X} \mathbf{v}_i\|_2}$. Define the first-order time difference operator $\mathbf{D}_n \in \mathbb{R}^{(K-1) \times K}$ as:

\begin{equation}
\mathbf{D}_n = \begin{bmatrix}
-1 & 1 & 0 & \cdots & 0 \\
0 & -1 & 1 & \cdots & 0 \\
\vdots & \vdots & \ddots & \ddots & \vdots \\
0 & 0 & \cdots & -1 & 1
\end{bmatrix}
\label{eq:Dn}  
\end{equation}

Then, the differential energy of the modal component $s_i(n)$ satisfies:
\begin{equation}
\sum_{n=1}^{K-1}|\nabla_n s_i(n)|^2 = c\|\mathbf{D}_n \mathbf{X}_i^T\|_F^2 = c \mathbf{v}_i^T \mathbf{R}_n \mathbf{v}_i \cdot \|\mathbf{u}_i\|_2^2
\label{eq:energy_sum}
\end{equation}

where $c > 0$ is a constant factor determined by the diagonal weights of the trajectory matrix, and when $L << K$, we have $c \approx \frac{1}{L}$.$\mathbf{R}_n = \mathbf{D}_n^T \mathbf{D}_n$ is a time-domain positive semi-definite matrix.

\subsection*{Proof:}

Consider the diagonal averaging property of the trajectory matrix constructed by Hankel's rule. For the global differential energy of the modal component $s_i(n)$:

\begin{equation}
\mathcal{E}_n = \sum_{n=1}^{N-1}|\nabla_n s_i(n)|^2 = c \|\mathbf{D}_n \mathbf{X}_i^T\|_F^2
\label{eq:global_energy}
\end{equation}

Expanding the Frobenius norm:

\begin{equation}
\|\mathbf{D}_t \mathbf{X}_i^T\|_F^2 = \mathrm{tr}\left( (\mathbf{D}_t \mathbf{X}_i^T)^T (\mathbf{D}_t \mathbf{X}_i^T) \right) = \mathrm{tr} \left( \mathbf{X}_i \mathbf{D}_t^T \mathbf{D}_t \mathbf{X}_i^T \right) = \mathrm{tr} \left( \mathbf{X}_i \mathbf{R}_t \mathbf{X}_i^T \right)
\label{eq:frobenius_expand}
\end{equation}

The trajectory matrix $\mathbf{X} \in \mathbb{R}^{L \times K}$ has a singular value decomposition (SVD) of:

\begin{equation}
\mathbf{X} = \mathbf{U} \Sigma \mathbf{V}^T = \sum_{i=1}^r \sigma_i \mathbf{u}_i \mathbf{v}_i^T
\label{eq:X_svd}
\end{equation}

where $\mathbf{u}_i$ is the left singular vector of $X^T X$, $\mathbf{v}_i$ is the right singular vector of $X X^T$, and $\sigma_i = \sqrt{\lambda_i}$ is the singular value (with $\lambda_i$ being the eigenvalue of $X^T X$ or $X X^T$).

Thus, we have:

\begin{equation}
\mathbf{X}_i^T = (\mathbf{u}_i \mathbf{v}_i^T)^T = \mathbf{v}_i \mathbf{u}_i^T \in \mathbb{R}^{K \times L}
\label{eq:XiT}
\end{equation}

Substituting into \eqref{eq:frobenius_expand}:

\begin{equation}
\mathrm{tr} \left( \mathbf{X}_i \mathbf{R}_n \mathbf{X}_i^T \right) = \mathrm{tr} \left( (\mathbf{u}_i \mathbf{v}_i^T) \mathbf{R}_n (\mathbf{v}_i \mathbf{u}_i^T) \right) = (\mathbf{v}_i^T \mathbf{R}_n \mathbf{v}_i) \cdot \|\mathbf{u}_i\|_2^2
\label{eq:tr_expand}
\end{equation}

Thus, we obtain:

\begin{equation}
c \|\mathbf{D}_n \mathbf{X}_i^T\|_F^2 = c \mathbf{v}_i^T \mathbf{R}_n \mathbf{v}_i \cdot \|\mathbf{u}_i\|_2^2
\label{eq:final_energy}
\end{equation}

This completes the proof.

Furthermore:

\begin{equation}
\mathbf{D}_n \mathbf{X}_i^T = \mathbf{D}_n (\mathbf{v}_i \mathbf{u}_i^T) = (\mathbf{D}_n \mathbf{v}_i) \mathbf{u}_i^T \in \mathbb{R}^{(K-1) \times L}
\label{eq:DnXiT}
\end{equation}

Define $\mathbf{w}_i = \mathbf{D}_n \mathbf{v}_i \in \mathbb{R}^{K-1}$, then:

\begin{equation}
\mathbf{D}_n \mathbf{X}_i^T = \mathbf{w}_i \mathbf{u}_i^T = \begin{bmatrix}
w_{i1} \mathbf{u}_i^T \\
w_{i2} \mathbf{u}_i^T \\
\vdots \\
w_{i(K-1)} \mathbf{u}_i^T
\end{bmatrix}
\label{eq:DnXiT_w}
\end{equation}

where $\mathbf{w}_i$ represents the time difference result:

\begin{equation}
w_{ik} = v_{i(k+1)} - v_{ik}, \quad k = 1, \ldots, K-1
\label{eq:w_ik}
\end{equation}

Substituting into \eqref{eq:global_energy}, we get:

\begin{equation}
\sum_{t=1}^{N-1}|\nabla_t s_i(t)|^2 = c \|\mathbf{D}_n \mathbf{X}_i^T\|_F^2 = c \|\mathbf{w}_i\|_2^2 \|\mathbf{u}_i\|_2^2
\label{eq:energy_w}
\end{equation}

Next, we construct the regularization objective function. Mode separation is equivalent to maximizing the variance of the Gram matrix, but the proposed method also minimizes the mode bandwidth. Therefore, we define the objective function of the Gram matrix as:

\begin{equation}
\mathcal{J}(\mathbf{w}) = \underbrace{\mathbf{w}^T \mathbf{G} \mathbf{w}}_{\text{principal component variance}} - \mu \underbrace{\mathbf{w}^T \mathbf{X}^T \mathbf{\Omega} \mathbf{X} \mathbf{w}}_{\text{differential energy}}
\label{eq:objective}
\end{equation}

The first term on the right-hand side is the variance term, and the second term is the bandwidth term (equivalent to $s(t) = \sum_i u_i(t)$). Here, $\mathbf{Q} = \mathbf{X}^T \mathbf{R}_t \mathbf{X}$.

Similarly, to achieve greater smoothing, a second-order time difference operator can be chosen:

\begin{equation}
\mathbf{D}_2 = \begin{bmatrix}
1 & -2 & 1 & 0 & \cdots \\
0 & 1 & -2 & 1 & \cdots \\
\vdots & \vdots & \vdots & \vdots & \ddots
\end{bmatrix}
\label{eq:D2}
\end{equation}

\subsection*{Theorem 2:}

Maximizing $\mathcal{J}(\mathbf{v})$ is equivalent to solving the generalized eigenvalue problem:

\begin{equation}
\mathbf{G} \mathbf{v} = \gamma (\mathbf{I} + \alpha \mathbf{Q}) \mathbf{v}, \quad \alpha = \frac{\beta}{\gamma}
\label{eq:general_eigen}
\end{equation}

where $\gamma$ is the generalized eigenvalue and $\mathbf{v}$ is the corresponding eigenvector.

\subsection*{Proof:}

We construct the Lagrangian function and establish the constrained optimization problem:

\begin{equation}
\mathcal{L}(\mathbf{v}, \gamma) = \mathbf{v}^T \mathbf{G} \mathbf{v} - \beta \cdot \mathbf{v}^T \mathbf{Q} \mathbf{v} - \gamma (\mathbf{v}^T \mathbf{v} - 1)
\label{eq:lagrangian}
\end{equation}

Due to the constraint \( \mathbf{v}^{T} \mathbf{v} = 1 \), which defines a compact set (the unit sphere), the function \( \mathcal{L} \) is convex. Therefore, the condition for the extremum is:

\begin{equation}
\nabla_\mathbf{v} \mathcal{L} = 2 \mathbf{G} \mathbf{v} - 2 \beta \mathbf{Q} \mathbf{v} - 2 \gamma \mathbf{v} = 0
\label{eq:extremum_condition}
\end{equation}
Simplifying, we obtain:

\begin{equation}
\mathbf{Gv} = \gamma \mathbf{v} + \beta \mathbf{Qv}
\label{eq:simplified_condition}
\end{equation}

Defining \( \alpha = \frac{\beta}{\gamma} \), we get:

\begin{equation}
\mathbf{Gv} = \gamma (\mathbf{I} + \alpha \mathbf{Q}) \mathbf{v} = \gamma \mathbf{M} \mathbf{v}
\label{eq:generalized_eigen_final}
\end{equation}

(1) \textbf{Boundary Conditions:}

- When \( \beta \to 0 \), this degenerates to the standard PCA problem \( \mathbf{Gv} = \gamma \mathbf{v} \).
- When \( \beta \to \infty \), we obtain the limiting smooth solution.

(2) \textbf{Properties of the Generalized Eigenvalue Problem:}

- The matrix \( \mathbf{M} = \mathbf{I} + \alpha \mathbf{Q} \) is positive definite (\( \mathbf{Q} \geq 0, \alpha > 0 \)).
- The eigenvectors satisfy \( \mathbf{v}_{i}^{T} (\mathbf{I} + \alpha \mathbf{Q}) \mathbf{v}_{j} = \delta_{ij} \).

Additionally, the eigenvectors \( \mathbf{v}_i \) are orthogonal in the inner product space defined by \( \mathbf{M} \), i.e., \( M \)-orthogonality. This can be proven as follows:

Let \( \mathbf{v}_i \) and \( \mathbf{v}_j \) be solutions to the generalized eigenvalue problem, with corresponding eigenvalues \( \gamma_i \) and \( \gamma_j \). Then:

\begin{equation}
\mathbf{G} \mathbf{v}_i = \gamma_i \mathbf{M} \mathbf{v}_i, \quad \mathbf{G} \mathbf{v}_j = \gamma_j \mathbf{M} \mathbf{v}_j
\label{eq:eigenvector_pairs}
\end{equation}

Left-multiply both equations by \( \mathbf{v}_j^T \) and \( \mathbf{v}_i^T \) respectively:

\begin{equation}
\mathbf{v}_j^T \mathbf{G} \mathbf{v}_i = \gamma_i \mathbf{v}_j^T \mathbf{M} \mathbf{v}_i, \quad \mathbf{v}_i^T \mathbf{G} \mathbf{v}_j = \gamma_j \mathbf{v}_i^T \mathbf{M} \mathbf{v}_j
\label{eq:left_multiplied}
\end{equation}

Since both \( \mathbf{G} \) and \( \mathbf{M} \) are symmetric matrices, we have:

\begin{equation}
\mathbf{v}_j^T \mathbf{G} \mathbf{v}_i = \mathbf{v}_i^T \mathbf{G} \mathbf{v}_j, \quad \mathbf{v}_j^T \mathbf{M} \mathbf{v}_i = \mathbf{v}_i^T \mathbf{M} \mathbf{v}_j
\label{eq:symmetry_properties}
\end{equation}

Now, combining the equations:

\begin{equation}
(\gamma_i - \gamma_j) \mathbf{v}_i^T \mathbf{M} \mathbf{v}_j = 0
\label{eq:orthogonality_condition}
\end{equation}

If \( \gamma_i \neq \gamma_j \), then \( \mathbf{v}_i^T \mathbf{M} \mathbf{v}_j = 0 \) (orthogonality).

If \( i = j \), we can normalize so that \( \mathbf{v}_i^T \mathbf{M} \mathbf{v}_i = 1 \).
\subsection{Anti-noise Performance Analysis}
Given the observed signal:
\begin{equation}
x(t) = s(t) + n(t)
\label{eq:observed_signal}
\end{equation}
where \( s(t) \) is the true signal, and \( n(t) \sim \mathcal{N}(0, \sigma_n^2) \) is the additive Gaussian white noise.

We construct the trajectory matrix based on Hankel's rule:
\begin{equation}
\mathbf{X} = \mathbf{S} + \mathbf{N} \in \mathbb{R}^{L \times K}
\label{eq:trajectory_matrix}
\end{equation}

Now, we define the Gram matrix:
\begin{equation}
\mathbf{G} = \mathbf{X}^T \mathbf{X} = \mathbf{S}^T \mathbf{S} + \mathbf{S}^T \mathbf{N} + \mathbf{N}^T \mathbf{S} + \mathbf{N}^T \mathbf{N} = \mathbf{A} + \mathbf{E}
\label{eq:gram_matrix}
\end{equation}
where \( \mathbf{A} = \mathbf{S}^T \mathbf{S} \) is the Gram matrix of the true signal, and
\begin{equation}
\mathbf{E} = \mathbf{S}^T \mathbf{N} + \mathbf{N}^T \mathbf{S} + \mathbf{N}^T \mathbf{N}
\label{eq:perturbation_matrix}
\end{equation}
is the perturbation introduced by the noise.

The cross-term expectation is:
\begin{equation}
\mathbb{E}[\mathbf{S}^T \mathbf{N}] = \mathbf{0}
\label{eq:cross_term_expectation}
\end{equation}

The noise term expectation is:
\begin{equation}
\mathbb{E}[\mathbf{N}^T \mathbf{N}] = \sigma_n^2 L \mathbf{I}_K
\label{eq:noise_term_expectation}
\end{equation}

The noise power is:
\begin{equation}
\|\mathbf{N}^T \mathbf{N}\|_F^2 = \sum_{i,j} |n_i^T n_j|^2 \approx KL \sigma_n^4
\label{eq:noise_power}
\end{equation}
where \( \|\cdot\| \) represents the spectral norm (the largest singular value of the matrix).

By using Weyl's inequality:
\begin{equation}
|\lambda_i(\mathbf{G}) - \lambda_i(\mathbf{S}^T \mathbf{S})| \leq \|\mathbf{E}\|
\label{eq:weyl_inequality}
\end{equation}

Also, since:
\begin{equation}
\|\mathbf{E}\| \leq \|\mathbf{S}\| \|\mathbf{N}\| + \|\mathbf{S}\| \|\mathbf{N}\| + \|\mathbf{N}\|^2 = 2 \|\mathbf{S}\| \|\mathbf{N}\| + \|\mathbf{N}\|^2
\label{eq:e_norm_bound}
\end{equation}
we obtain the upper bound of the eigenvalue perturbation:
\begin{equation}
|\lambda_i(\mathbf{G}) - \lambda_i(\mathbf{S}^T \mathbf{S})| \leq 2 \|\mathbf{S}\| \|\mathbf{N}\| + \|\mathbf{N}\|^2
\label{eq:eigenvalue_perturbation_bound}
\end{equation}

The proposed method solves the generalized eigenvalue problem:
\begin{equation}
\mathbf{Gv} = \gamma (\mathbf{I} + \alpha \mathbf{Q}) \mathbf{v} = \gamma \mathbf{M} \mathbf{v}
\label{eq:generalized_eigenproblem}
\end{equation}

Eigenvalue correction:
\begin{equation}
\gamma_i = \frac{\lambda_i(\mathbf{G})}{1 + \alpha \mu_i}, \quad \mu_i = \mathbf{v}_i^T \mathbf{Q} \mathbf{v}_i
\label{eq:eigenvalue_correction}
\end{equation}
where \( \mu_i \) represents the quadratic form of the \(i\)-th mode under the regularization matrix \( \mathbf{Q} \), which constrains the bandwidth of that mode.

In general, Gaussian noise is broadband, and from the above equation, we see that when \( \mu_i \) is large, wideband noise is suppressed, whereas narrowband signals are preserved.

Thus, the eigenvalue perturbation upper bound is corrected as:
\begin{equation}
|\gamma_i - \lambda_i(\mathbf{S}^T \mathbf{S})| \leq \frac{2 \|\mathbf{S}\|_2 \|\mathbf{N}\|_2 + \|\mathbf{N}\|_2^2}{1 + \alpha \mu_{\min}}
\label{eq:corrected_perturbation_bound}
\end{equation}

Therefore, eigenvalue perturbations can be compressed.

Next, consider the reconstruction error analysis. Suppose the true signal is \( \mathbf{s} \), and the signal reconstructed from the noisy signal using the regularized modes is \( \hat{\mathbf{s}} \):
\begin{equation}
\hat{\mathbf{s}} = \sum_{i=1}^r g_i \mathbf{X} \mathbf{v}_i \mathbf{v}_i^T, \quad g_i = \frac{1}{1 + \alpha \mu_i}
\label{eq:reconstructed_signal}
\end{equation}
where \( r \) is the number of reconstructed modes (\( r \leq d \)).

The mean square error (MSE) is:
\begin{equation}
\mathcal{E} = \|\mathbf{s} - \mathbb{E}[\hat{\mathbf{s}}]\|^2 + \mathbb{E} \|\hat{\mathbf{s}} - \mathbb{E}[\hat{\mathbf{s}}]\|^2
\label{eq:mse}
\end{equation}
where the first term is the bias (squared) and the second term is the variance.

For the bias (squared) term, since the modes are orthogonal:
\begin{equation}
\mathrm{Bias}^2 = \|\mathbf{s} - \mathbb{E}[\hat{\mathbf{s}}]\|^2 = \left\|\sum_{i=r+1}^d \mathbf{s}_i + \sum_{i=1}^r (1 - g_i) \mathbf{s}_i \right\|^2 = \sum_{i=1}^r \left(\frac{\alpha \mu_i}{1 + \alpha \mu_i}\right)^2 \|\mathbf{s}_i\|^2 + \sum_{i=r+1}^d \|\mathbf{s}_i\|^2
\label{eq:bias_squared}
\end{equation}

For standard PCA:
\begin{equation}
\mathrm{Bias}_{\mathrm{PCA}}^2 = \|\mathbf{s} - \mathbb{E}[\hat{\mathbf{s}}_{\mathrm{PCA}}]\|^2 \approx \sum_{i=r+1}^d \|\mathbf{s}_i\|^2
\label{eq:pca_bias}
\end{equation}

In comparison, the proposed method introduces an additional term, indicating the energy of the signal that is overly attenuated due to regularization. Fortunately, for narrowband signals, this term is small, while for wideband noise, this term is larger.

For the variance term, standard PCA gives:
\begin{equation}
\mathrm{Variance}_{\mathrm{PCA}} = \mathbb{E} \left[ \|\hat{\mathbf{s}}_{\mathrm{PCA}} - \mathbb{E}[\hat{\mathbf{s}}_{\mathrm{PCA}}]\|^2 \right] = \sigma_n^2 L r
\label{eq:pca_variance}
\end{equation}

The variance term for the proposed method is:
\begin{equation}
\mathrm{Variance}_{\mathrm{reg}} = \sigma_n^2 L \sum_{i=1}^r \left( \frac{1}{1 + \alpha \mu_i} \right)^2
\label{eq:reg_variance}
\end{equation}

Thus, the ratio of the variance is:
\begin{equation}
\frac{\mathrm{Variance}_{\mathrm{reg}}}{\mathrm{Variance}_{\mathrm{PCA}}} = \frac{1}{r} \sum_{i=1}^r \frac{1}{(1 + \alpha \mu_i)^2} \leq 1
\label{eq:variance_ratio}
\end{equation}

This shows that the variance is significantly reduced.

In conclusion, although the proposed method increases the bias term, it mainly affects the noise modes, while significantly reducing the variance. By selecting an appropriate regularization parameter \( \alpha \), a good balance can be achieved between bias and variance, which is especially useful in scenarios with low signal-to-noise ratio (SNR) and high-frequency noise.
\section {Robust Mode Decomposition (RMD) Method Overview}
In this section, we introduce the proposed RMD method, which is primarily based on the concepts discussed in the previous section. The goal of RMD is to construct an augmented matrix such that the separated signals preserve the physical characteristics of the original signals, avoid separating spurious modes, and constrain the bandwidth of the modal signals to improve noise and interference robustness.

Based on the principles proposed in Section II, RMD can be divided into four main steps. The first step is to map the time series to a phase space trajectory matrix. This process involves two parameters: embedding dimension \(K\) and delay \(\tau\). Based on empirical experience, \(\tau\) is generally set to 1, and \(K\) is set to one-third of the time series length or determined according to the spectral peak of the time series. Specifically, it is determined by the frequency corresponding to the maximum peak in the power spectral density (PSD) of the time series\cite{24}:
\begin{equation}
K = 1.2 \times \frac{f_{\text{max}}}{F_s}
\label{eq:embedding_dim}
\end{equation}

In general, when the value of the above formula is small (threshold set to \(10^3\) in this paper), it indicates that the dominant frequency of the time series is low, and the embedding dimension should be increased to improve frequency resolution. Thus, \(K\) is set to one-third of the sequence length. Conversely, if the signal frequency is high, a smaller embedding dimension can capture the local dynamics. It is important to note that at low signal-to-noise ratios (SNR), the adaptive rule described above may have limited effect, and a larger embedding dimension is generally preferred.

The second step is to compute the Gram matrix \(\mathbf{G} = \mathbf{X}^T \mathbf{X}\) from the trajectory matrix \(\mathbf{X}\), and then select the regularization factor. If a time-domain difference operator is used to construct the augmented matrix, we solve the generalized eigenvalue problem in \eqref{eq:generalized_eigenproblem}

The third step is to arrange the obtained eigenvalues in descending order, and then, based on the user-specified number of modal components, cluster and merge the eigenvectors corresponding to larger eigenvalues to eliminate the influence of multiple roots. At the same time, eigenvectors corresponding to smaller eigenvalues are discarded as high-frequency noise. Since the dimension of eigenvectors is much smaller than the dimension of the signal, the computational load is much smaller after the recovery process. The method is as follows: first, select the eigenvector corresponding to the largest eigenvalue, then traverse the next few eigenvectors, merging those with high similarity (either select the first eigenvector or directly add them), and then delete the merged eigenvectors. This process is repeated from the residuals. The decomposition stops when the required number of modes is reached or when all eigenvectors have been used, returning each mode's corresponding eigenvector and the residual eigenvector.
\begin{equation}
\mathbf{v} = \sum_{i=1}^k \mathbf{v}_i + \mathbf{v}_r
\label{eq:eigenvector_sum}
\end{equation}
where \(k\) is the given number of modes or the maximum number of modes supported, \(\mathbf{v}_i\) is the eigenvector corresponding to the \(i\)-th mode, and \(\mathbf{v}_r\) is the residual eigenvector.

The similarity measure can be chosen from the following three options or their combination, and it can also be extended to any other measure:

- Cosine similarity \cite{25}:
\begin{equation}
\text{sim}(\mathbf{v}_i, \mathbf{v}_j) = \frac{\mathbf{v}_i \cdot \mathbf{v}_j}{\|\mathbf{v}_i\| \cdot \|\mathbf{v}_j\|}
\label{eq:cosine_sim}
\end{equation}

- Pearson correlation coefficient\cite{26}:
\begin{equation}
r(\mathbf{v}_i, \mathbf{v}_j) = \frac{\sum (\mathbf{v}_i - \bar{\mathbf{v}}_i)(\mathbf{v}_j - \bar{\mathbf{v}}_j)}{\sigma_{\mathbf{v}_i} \sigma_{\mathbf{v}_j}}
\label{eq:pearson_corr}
\end{equation}

- Normalized Euclidean distance\cite{27}:
\begin{equation}
d_{\text{norm}}(\mathbf{v}_i, \mathbf{v}_j) = \sqrt{\sum_{k=1}^n \frac{(v_{i,k} - v_{j,k})^2}{\sigma_k^2}}
\label{eq:normalized_ed}
\end{equation}

The final step is to reconstruct the trajectory matrix using the eigenvectors obtained in step 3:
\begin{equation}
\mathbf{Z} = \sum_{i=1}^k \mathbf{Z}_i + \mathbf{Z}_r
\label{eq:trajectory_recon}
\end{equation}
where \(\mathbf{Z}_i = \mathbf{V}_r^T \mathbf{X}\). The separated modal components are then obtained by diagonal averaging:
\begin{equation}
\mathbf{z}(n) = \sum_{i=1}^k \mathbf{z}_i(n) + \mathbf{z}_r(n)
\label{eq:modal_recon}
\end{equation}

The  pseudocode for the above steps are as follows:

\begin{algorithm}[H]  
\caption{Robust Mode Decomposition (RMD)}
\begin{algorithmic}[1]
\REQUIRE $x$: Input signal of length $N$
\REQUIRE $r$: Preset number of modes
\REQUIRE $\theta$: Similarity merging threshold (default $0.85$)
\REQUIRE $\alpha$: Regularization factor (default $0.3$)
\ENSURE $V_{\text{main}}$: Main mode eigenvectors $[v_1, v_2, ..., v_r]$
\ENSURE $V_{\text{res}}$: Residual eigenvectors
\ENSURE $z$: Reconstructed modal components $[z_1, z_2, ..., z_r, z_r]$

\STATE \textbf{Step 1: Trajectory Matrix Construction}
\STATE Set delay parameter $\tau \leftarrow 1$
\STATE Find the maximum peak frequency $f_{\text{max}} \leftarrow \text{argmax}(PSD(x))$
\IF{$f_{\text{max}} / F_s < 1e-3$}
    \STATE Set embedding dimension $K \leftarrow N / 3$
\ELSE
    \STATE Set $K \leftarrow 1.2 \times F_s / f_{\text{max}}$
\ENDIF
\STATE Construct trajectory matrix $\mathbf{X} \leftarrow \text{Hankel}(x, L=K, \tau=\tau)$

\STATE \textbf{Step 2: Gram Matrix and Regularization}
\STATE Compute the Gram matrix $\mathbf{G} \leftarrow \mathbf{X}^T \mathbf{X}$
\STATE Construct first-order difference operator $D_n \leftarrow \text{construct\_1st\_diff\_matrix}(K)$
\STATE Compute smoothing matrix $R_n \leftarrow D_n^T D_n$
\STATE Set augmented matrix $\mathbf{M} \leftarrow \mathbf{I} + \alpha R_n$

\STATE \textbf{Step 3: Solve Generalized Eigenvalue Problem}
\STATE Solve generalized eigenvalue problem $\mathbf{G} \mathbf{v} = \gamma \mathbf{M} \mathbf{v}$
\STATE Sort eigenvalues $\gamma_i$ and eigenvectors $\mathbf{v}_i$ in descending order

\STATE \textbf{Step 4: Eigenvector Clustering and Merging}
\STATE Initialize $V_{\text{main}} \leftarrow []$
\FOR{$i = 1$ to $K$}
    \IF{size of $V_{\text{main}}$ exceeds $r$}
        \STATE Break
    \ENDIF
    \IF{vector $\mathbf{v}_i$ is merged}
        \STATE Continue
    \ENDIF
    \STATE Initialize cluster $ \leftarrow [\mathbf{v}_i]$
    \FOR{$j = i + 1$ to $K$}
        \STATE Compute similarity $\text{sim} \leftarrow f_{\text{similarity}}(\mathbf{v}_i, \mathbf{v}_j)$
        \IF{$\text{sim} > \theta$}
            \STATE Merge eigenvector $\mathbf{v}_j$ with cluster
            \STATE Mark $\mathbf{v}_j$ as merged
        \ENDIF
    \ENDFOR
    \STATE Compute merged vector $\mathbf{v}_{\text{merged}} \leftarrow \text{mean}(cluster, \text{weights}=\gamma_{\text{cluster}})$
    \STATE Append $\mathbf{v}_{\text{merged}}$ to $V_{\text{main}}$
\ENDFOR

\STATE \textbf{Step 5: Residual Extraction and Reconstruction}
\STATE Extract residual eigenvectors $V_{\text{res}} \leftarrow \{ \mathbf{v}_j \mid \mathbf{v}_j \text{ not merged} \}$
\FOR{each $\mathbf{v}_k$ in $V_{\text{main}}$}
    \STATE Compute projection coefficients $S_k \leftarrow \mathbf{v}_k^T \mathbf{X}$
    \STATE Compute mode trajectory matrix $\mathbf{Z}_k \leftarrow \mathbf{v}_k S_k^T$
    \STATE Compute mode component $z_k \leftarrow \text{diagonal\_average}(\mathbf{Z}_k)$
\ENDFOR
\STATE Compute residual component $z_{\text{res}} \leftarrow \text{diagonal\_average}(\mathbf{X} - \sum \mathbf{Z}_k)$

\RETURN $V_{\text{main}}, V_{\text{res}}, z$
\end{algorithmic}
\end{algorithm}

\vspace{1em}  

\section{RESULTS FOR COMPONENTS SEPARATION}
The effectiveness of the proposed method is thoroughly validated through both synthetic simulation datasets and real datasets. First, in the synthetic data experiments, we test the noise robustness of the proposed method. We begin by adding several sinusoidal sequences of different frequencies and amplitudes and then overlaying Gaussian white noise to achieve signal-to-noise ratios (SNR) of -15dB and -5dB. The proposed method is compared against two major competing methods: VMD and SGMD (when the regularization coefficient $\gamma = 0$, RMD is equivalent to SGMD). These two methods represent some of the most advanced levels of modal decomposition techniques in their respective categories. In the second set of experiments, we assess the performance of the proposed method in nonlinear signal processing. The signals in this experiment consist of several amplitude-modulated and phase-modulated waves, with noise added at an SNR of 0dB.

\subsection{Synthetic Data}
\subsubsection{Noise Robustness}
The univariate signal model used consists of four sinusoidal oscillatory components. The sampling frequency is 200Hz, and the sampling duration is 10 seconds. The frequencies of the three sinusoidal oscillations are 2Hz, 5Hz, and 19Hz, with amplitudes of 3, 0.5, and 4, respectively. After adding Gaussian white noise, the signal’s SNRs are -15dB and -5dB. When comparing the three methods, a parameter search is conducted, particularly for the comparison between SGMD and RMD, where the embedding dimension is fixed at $K = 200$. The results under the condition of SNR = -15 dB corresponding to the best performance in each experiment are presented IN fig. 2, and each method is configured to decompose the signal into 4 modes.
  \begin{figure}[htbp]
    \centering
    \includegraphics[width=1\linewidth]{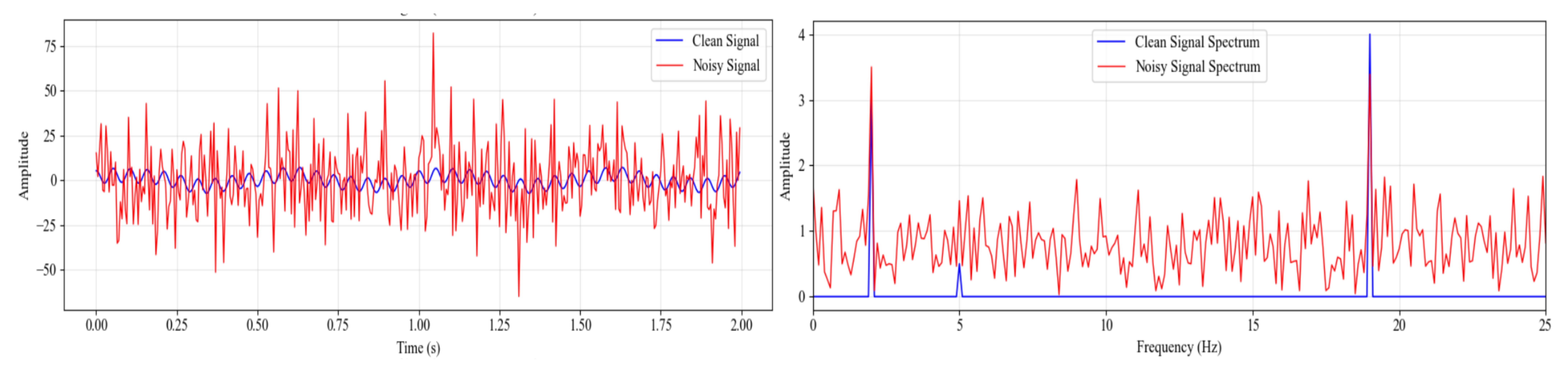}
    \caption{\small presents the time-domain waveforms and spectrograms of the signals used in simulations under the SNR of -15 dB. For better visualization, only the first 2 seconds of the time-domain waveforms are presented. The blue curves correspond to the original signals (before noise addition) and their spectrograms, while the red curves represent the signals after noise contamination and their respective spectrograms. As observed from the spectrograms, under the condition of -15 dB SNR, the signals suffer severe degradation—specifically, the 5 Hz signal component is almost completely submerged by noise.}
    \label{fig:2_signal_15dB}
  \end{figure}

\begin{figure}[H]  
    \centering
    \includegraphics[width=0.8\linewidth]{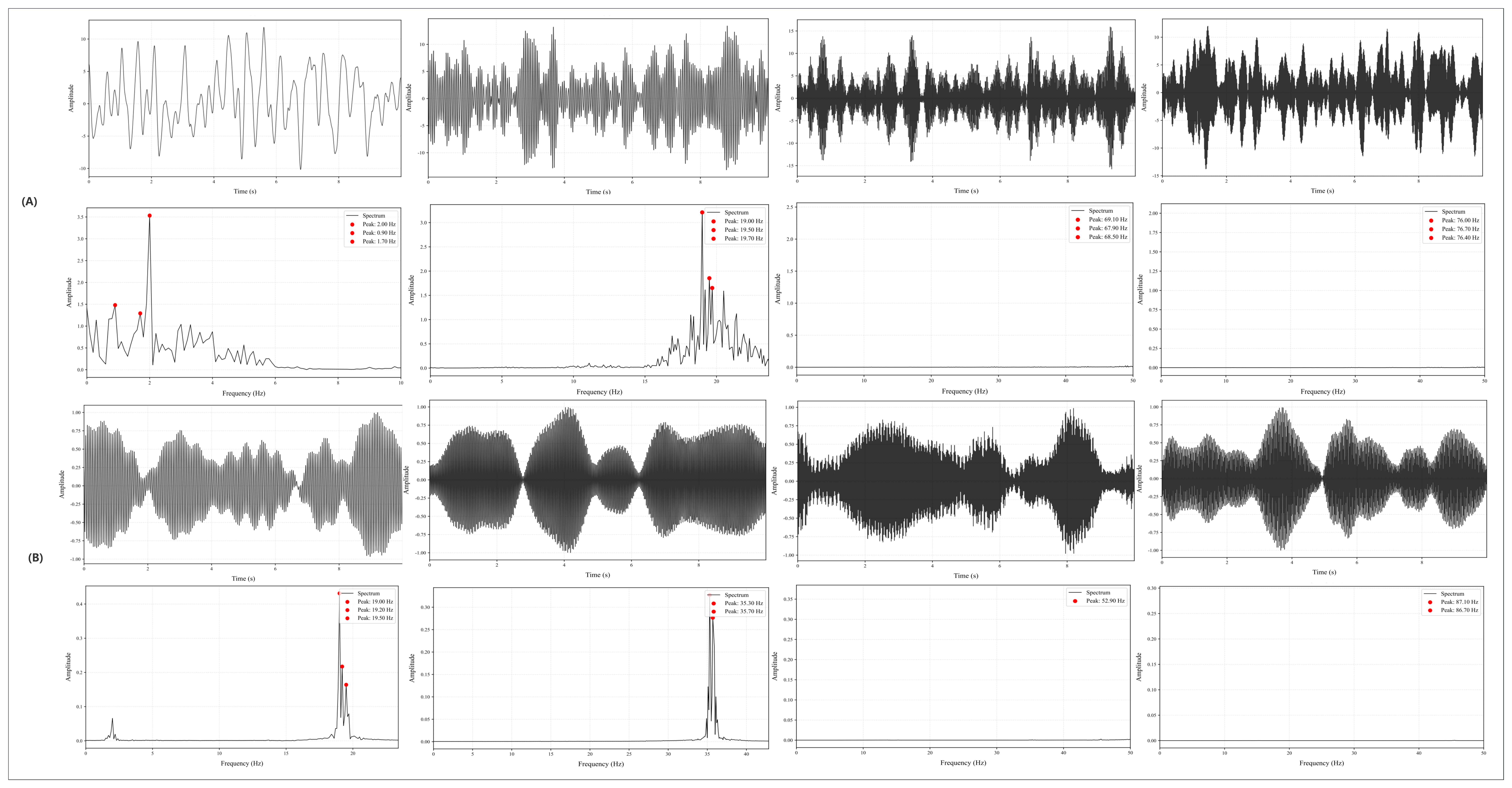}  
    \label{fig:signal_15dB_part1}  
\end{figure}

\clearpage  
\begin{figure}[H]  
    \centering
    \includegraphics[width=0.8\linewidth]{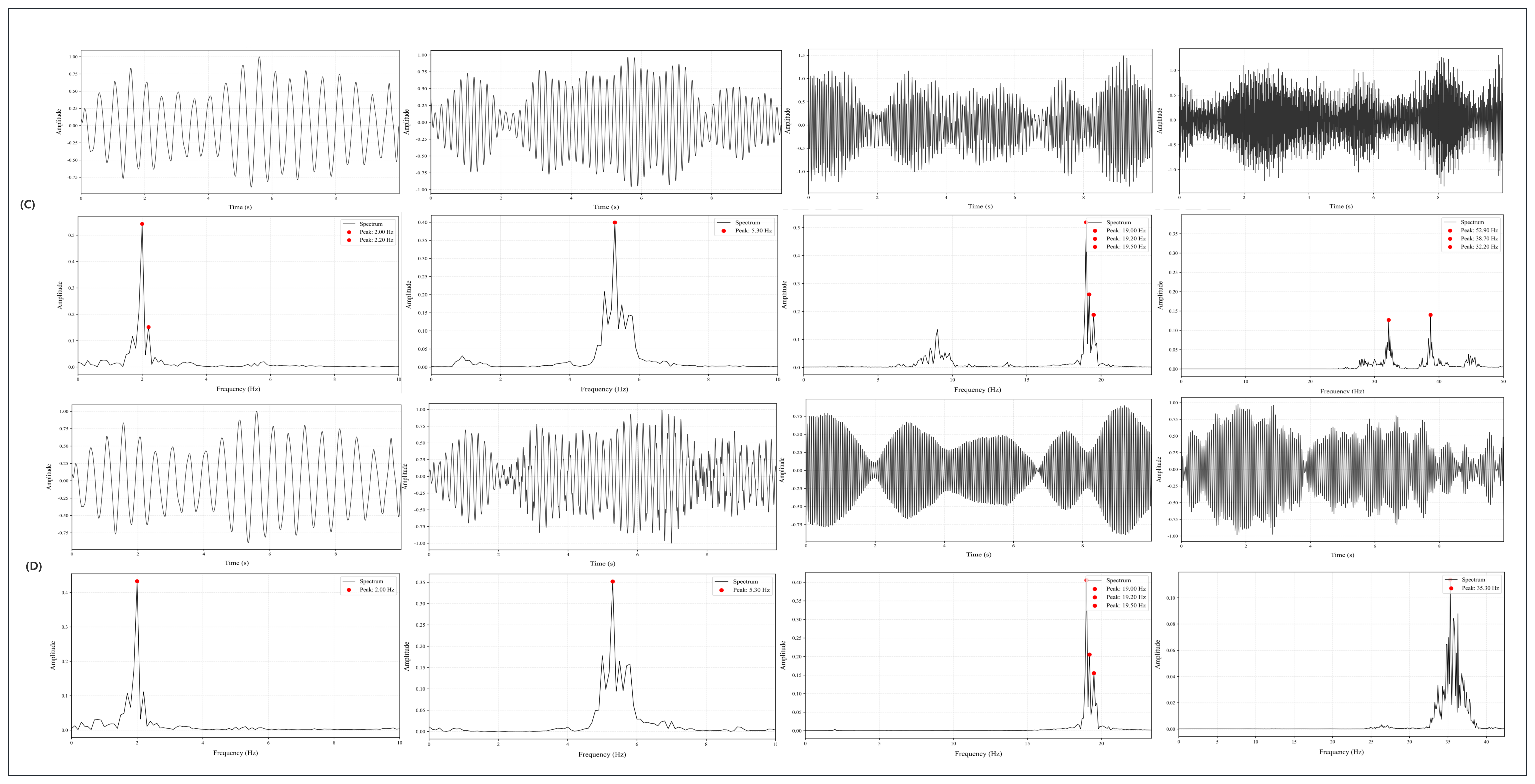}  
    \caption{\small (A) presents the time-domain waveforms and spectra of modes separated from the synthetic signal with SNR = -15 dB via VMD, and the spectral components beyond 50 Hz are not displayed. It can be observed that after adjusting the VMD parameters, VMD is capable of extracting the 2 Hz and 19 Hz components with relatively large amplitudes, but these components suffer from distortion. Additionally, the 5 Hz component with small amplitude is completely lost.(B) shows the processing results of the synthetic signal with SNR = -15 dB using SGMD. This result indicates that noise significantly degrades the processing performance of SGMD, enabling it to only separate the 19 Hz component.(C) and (D) depict the processing results of the synthetic signal with SNR = -15 dB using RMD under the first-order difference($\alpha = 10$) and second-order difference($\alpha = 0.15$) , respectively. These results demonstrate that under both conditions, RMD is the only method that can correctly separate all three modal components. Although the 5 Hz frequency band shifts due to the influence of noise, the overall modal component distortion is minor, which confirms the strong noise robustness of the proposed RMD.}
    \label{fig:signal_15dB_part2}  
\end{figure}
Moreover, we only adjusted the Signal-to-Noise Ratio (SNR) to -5 dB while keeping all other conditions unchanged, and repeated the experiment ,which is illustrated in Fig.4. The experimental results are illustrated in Fig.5.

  \begin{figure}[htbp]
    \centering
    \includegraphics[width=1\linewidth]{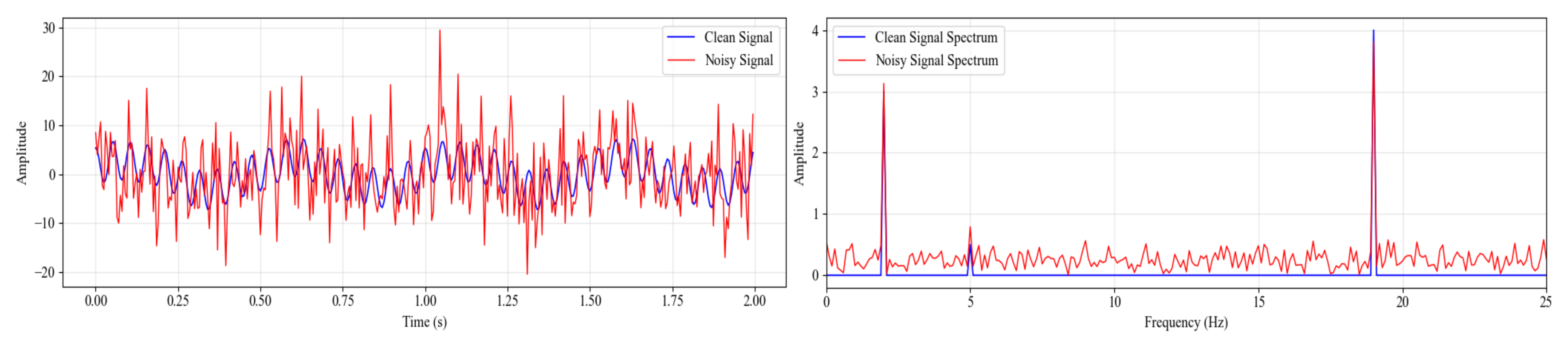}
    \caption{\small presents the time-domain waveforms and spectrograms of the signals used in simulations under an SNR of -5 dB. For enhanced clarity, only the first 2 seconds of the time-domain waveforms are shown. The blue curves correspond to the original signals (before noise addition) and their spectrograms, while the red curves represent the noise-contaminated signals and their respective spectrograms. As evident from the spectrograms, all components are visually identifiable under the -5 dB SNR condition.}
    \label{fig:4_signal_5dB}
  \end{figure}
  \begin{figure}[H]  
    \centering
    \includegraphics[width=0.8\linewidth]{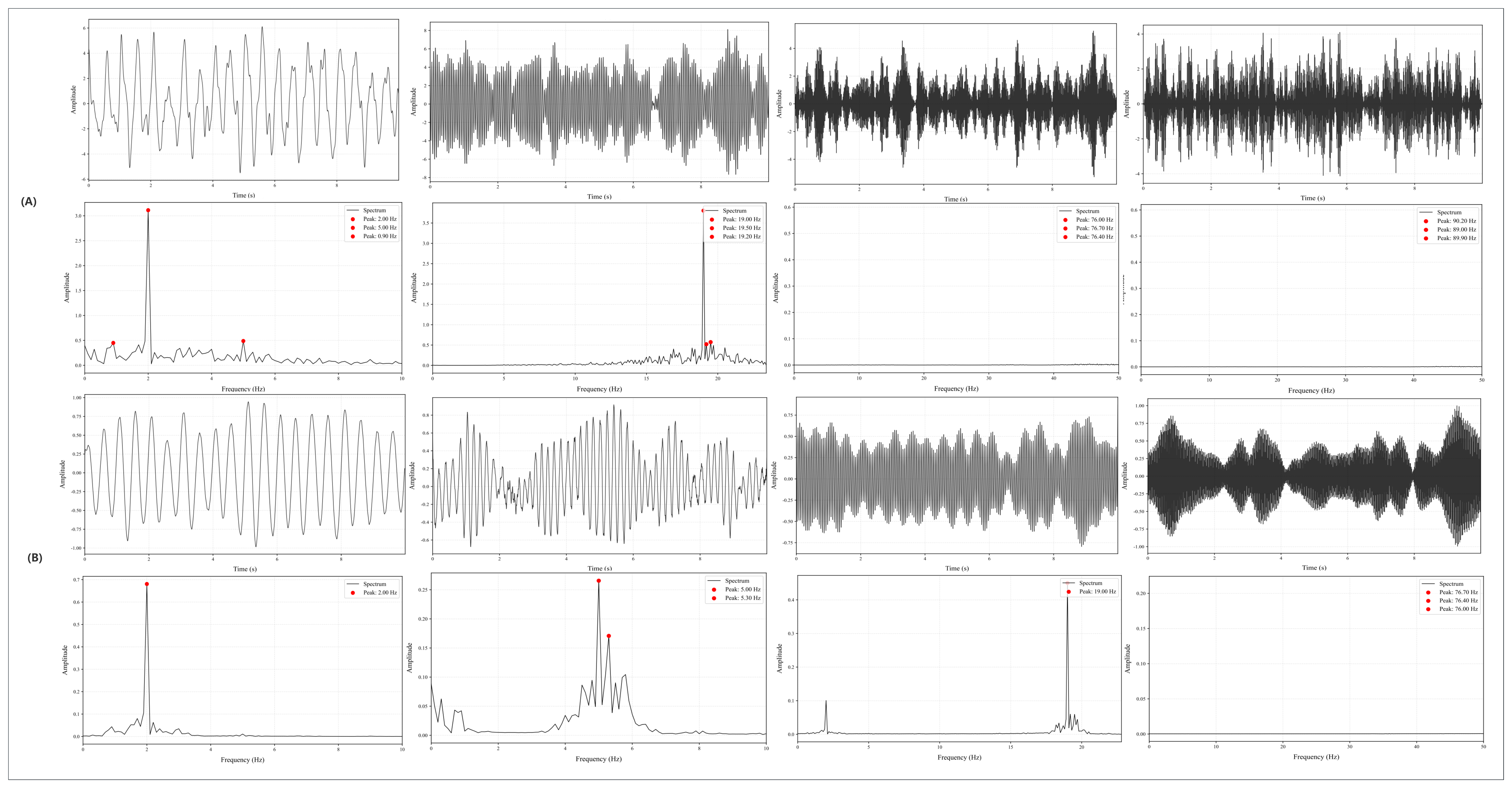}  
    \label{fig:signal_15dB_part1}  
\end{figure}
\begin{figure}[H]  
    \centering
    \includegraphics[width=0.8\linewidth]{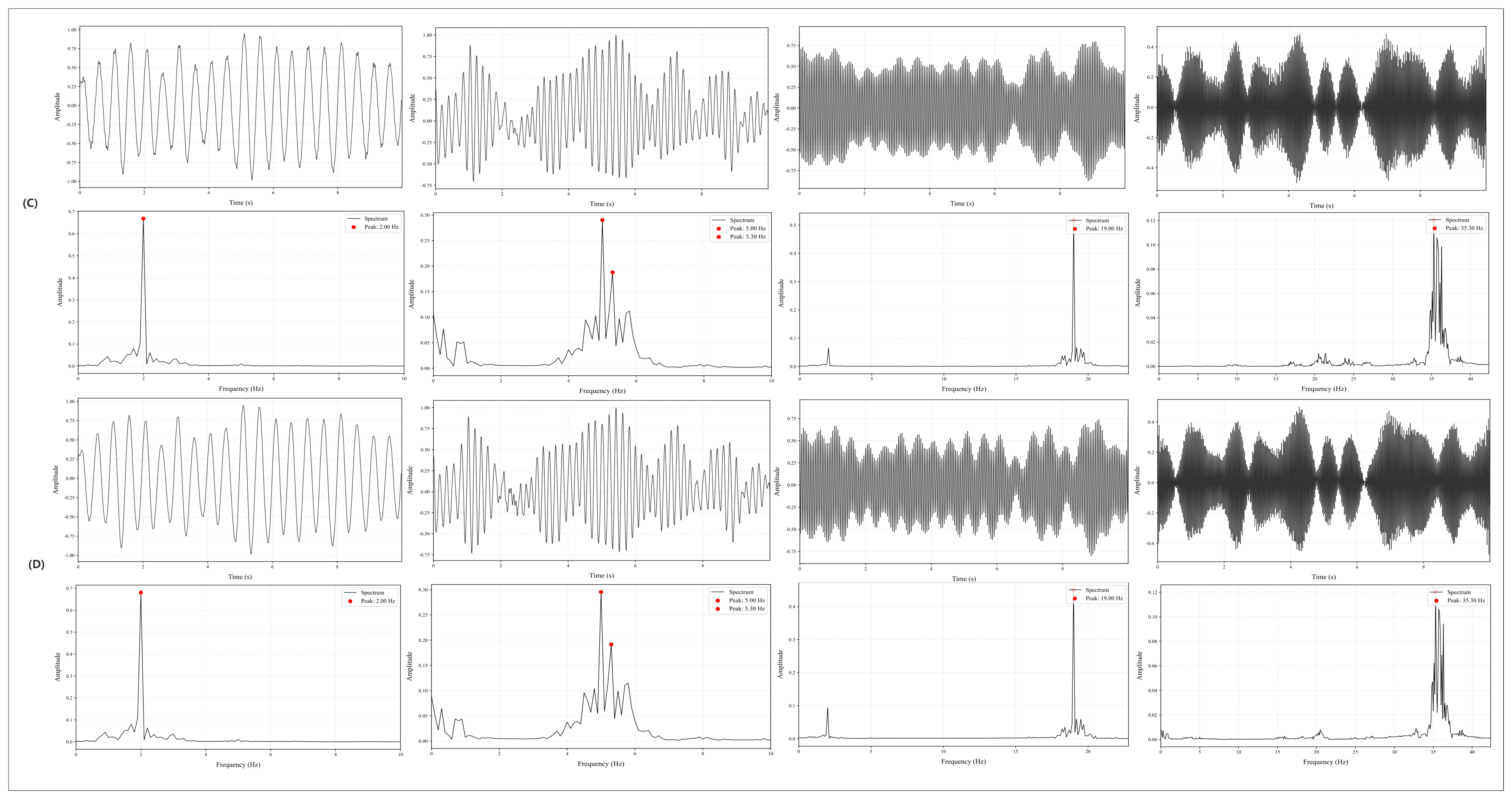}  
    \caption{\small (A) shows the decomposition results of VMD. Even after the SNR is improved, the 5 Hz component remains unseparated and is instead mixed with the 2 Hz component.(B), (C), and (D) present the decomposition results of SGMD, first-order RMD($\alpha = 8$), and second-order RMD($\alpha = 0.45$), respectively. As observed from the figures, all modal components are successfully separated, with only minor differences in amplitude.It can thus be concluded that compared to VMD, SGMD and RMD exhibit stronger noise robustness, as they adhere to the physical constraints of the intrinsic signal structure.}
    \label{fig:signal_15dB_part2}  
\end{figure}

\subsubsection{Nonlinear Signal Processing}

The noise-free combined signal in the univariate signal model is constructed by superposing three components, which is mathematically expressed as:$f = 2 \sin\left(2\pi f_1 t\right) \cdot \left[1 + 0.5 \sin\left(2\pi f_{\text{mod}} t\right)\right] + \sin\left(2\pi f_2 t\right) + \cos\left(2\pi f_3 t\right)$ where \( f_1 \), \( f_2 \), and \( f_3 \) denote the center frequencies of the three components, i.e., 3 Hz, 8 Hz, and 31 Hz, under the condition of SNR = 0 dB. All other settings remain the same as those specified in the summary of the previous subsection.
\begin{figure}[htbp]
    \centering
    \includegraphics[width=0.75\linewidth]{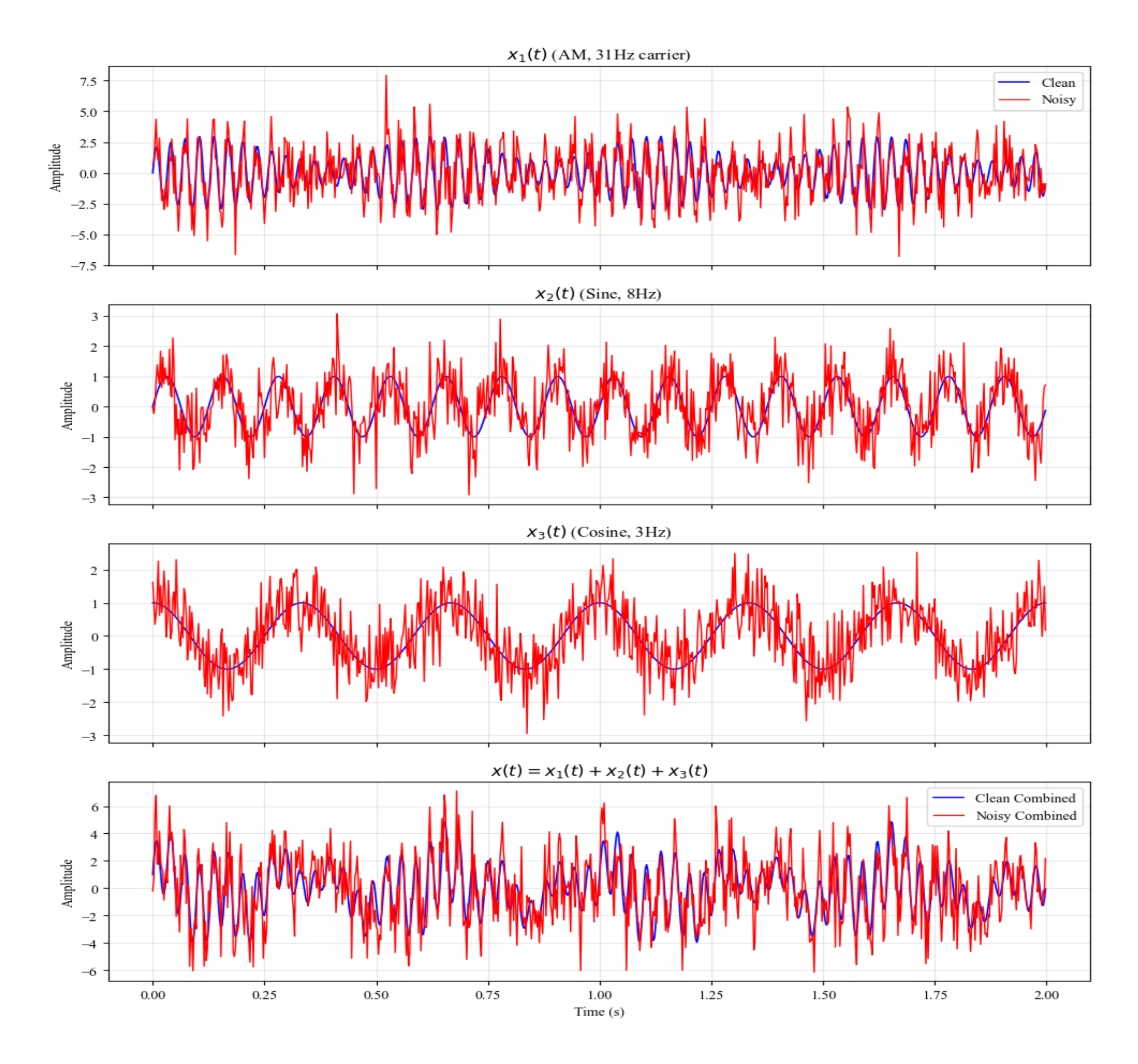}
    \caption{\small Presents the time-domain waveforms of the nonlinear signal to be processed in the experiment (with SNR = 0 dB), where the time-domain waveforms of the three components and the combined signal (first 2 seconds) are plotted. The blue curves represent the original signals, while the red ones correspond to the noise-added signals.}
    \label{fig:nonlinear_signal}
\end{figure}

\begin{figure}[H]
    \centering
    \includegraphics[width=0.8\linewidth]{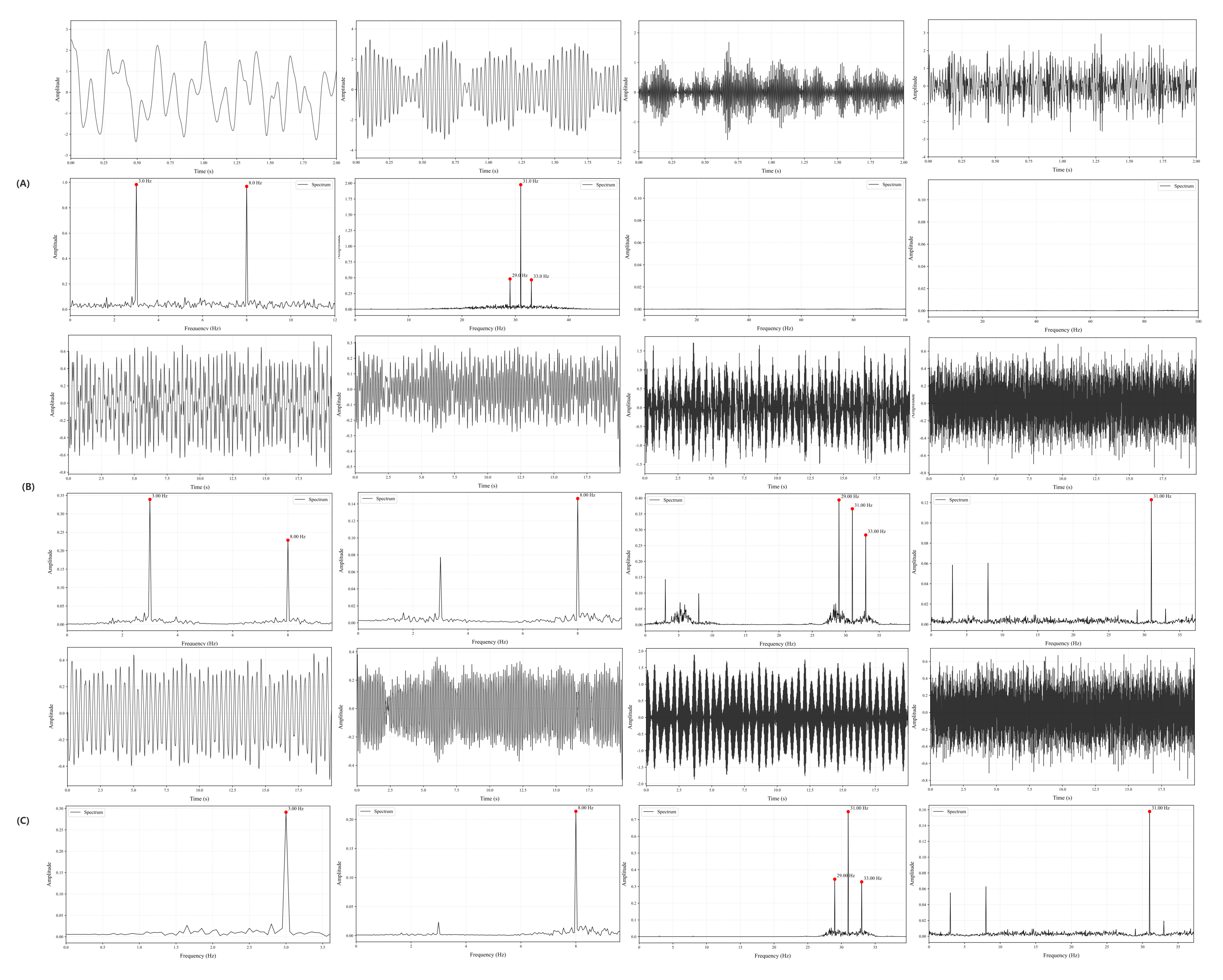}
    \caption{\small (A) shows VMD-processed modal waveforms and spectra. VMD poorly decomposes low frequencies (3 Hz and 8 Hz are fully aliased) but effectively decomposes the 31 Hz AM waveform. (B) presents SGMD results: poor decomposition with low-frequency aliasing and amplitude distortion in the 31 Hz AM waveform. (C) depicts first-order RMD results ($\alpha=2$; second-order RMD is also effective but omitted due to space). RMD achieves the best performance, correctly separating 3 Hz and 8 Hz (suppressing mutual interference) and preserving relative amplitudes in AM decomposition.}
    \label{fig:non_pro}
\end{figure}

\subsection{Real World Data Experiment}
In this subsection, the proposed method is applied to the decomposition of millimeter-wave radar signals. Under strong interference, the echo signals of millimeter-wave radar exhibit low signal-to-noise ratio (SNR) and contain numerous nonlinear components. The proposed method can adaptively decompose these signals to achieve denoising, laying a foundation for subsequent analysis. For comparison, VMD and SGMD are also tested in the experiment. 

The millimeter-wave radar dataset used herein was collected in our previous work \cite{28} using a TI AWR2243 millimeter-wave radar. This radar detects the micro-motion energy of the human chest, with a sampling frequency of 100 Hz and a total of 2048 sampling points. The goal of processing is to separate respiratory and heartbeat waveforms from the radar echo data, and the experimental results are shown in Fig. 8.
\begin{figure}[H]
    \centering
    \includegraphics[width=0.8\linewidth]{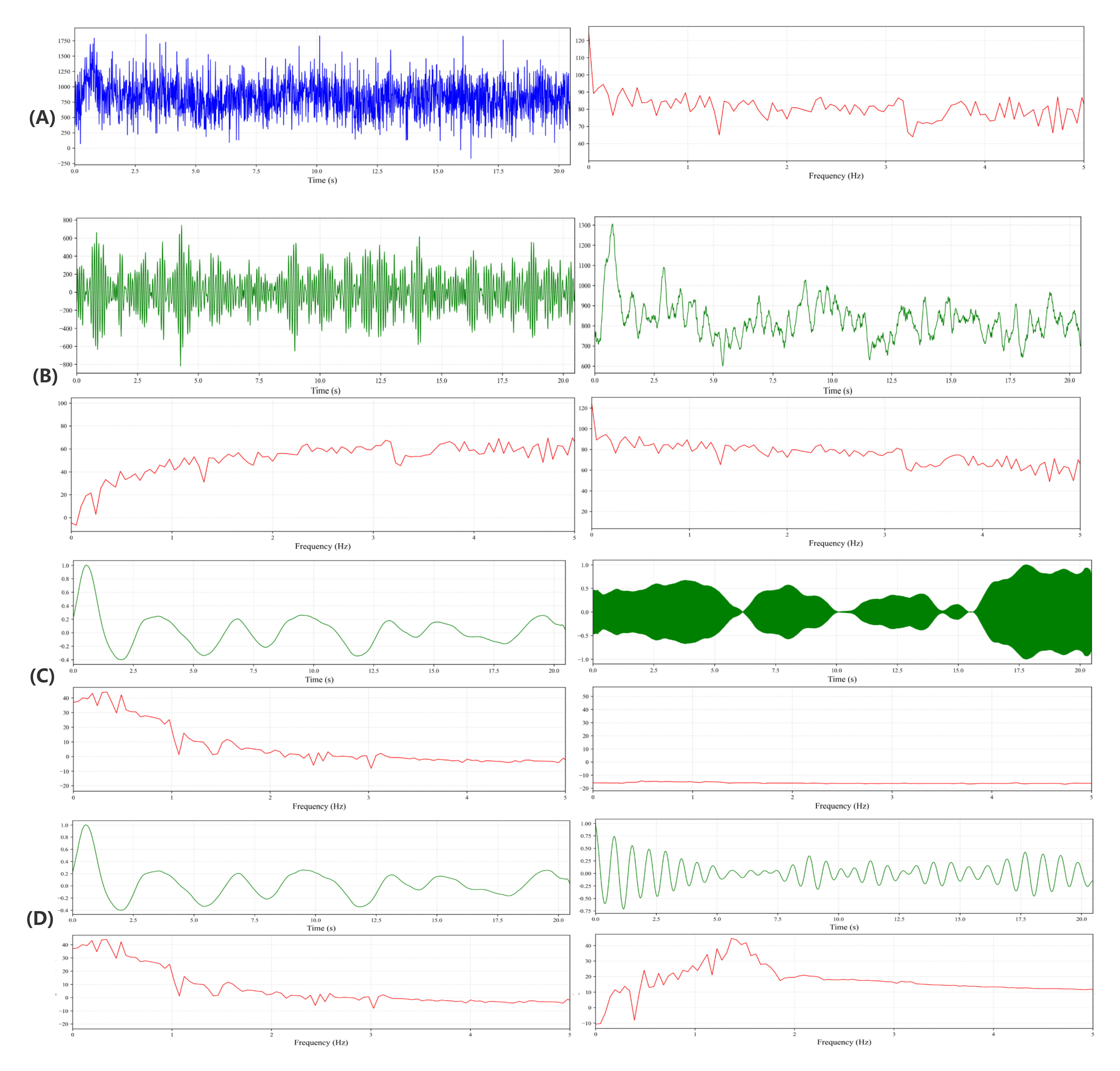}
    \caption{\small (A) shows the time-domain waveform and spectrum of micro-motion energy from radar echoes. Due to severe noise contamination, it is difficult to derive the trends of respiratory and heartbeat signals. (B) presents VMD processing results: no useful information is extracted because VMD exhibits weak performance in handling low-frequency signal energy and poor noise resistance. (C) depicts SGmd results: the respiratory waveform is extracted, but the heartbeat waveform is completely submerged in noise with spurious modes generated. (D) shows first-order RMD results ($\alpha=2$), where RMD correctly separates respiratory and heartbeat waveforms, and the measured frequencies are very close to those obtained by our ADS1293 ECG module.}
    \label{fig:radar}
\end{figure}
\section{Conclusion}
In this paper, we propose a novel mode decomposition method that combines the advantages of spectral analysis and variational optimization-based mode decomposition methods, which we call Robust Mode Decomposition (RMD). Inspired by the idea from VMD that limiting the signal bandwidth is equivalent to limiting the energy of the differential sequence, we innovatively introduce a time-domain differential operator during the eigendecomposition of the trajectory-GRAM matrix. This transforms the optimization problem into a generalized eigenvalue decomposition problem under the augmented matrix, thereby achieving a balance between bandwidth limitation and maintaining the intrinsic structure of the signal. This allows our method to avoid the generation of spurious modes to some extent and significantly enhances the noise robustness. We applied the proposed method to synthetic simulation data and real-world data, and the experiments demonstrate that our method outperforms spectral analysis methods such as SGMD in terms of noise robustness and nonlinear signal processing capability, while maintaining the intrinsic structure of the signal without producing spurious modes, as seen in iterative methods like VMD. The full code is available on GitHub.

Further research will focus on the following three aspects:

\begin{itemize}
    \item[(1)] This paper only considers real signals with low-pass bandwidth limitations. It is clear that the proposed method can easily be extended to the complex domain for the decomposition of analytic or IQ signals. Additionally, scenarios with band-pass bandwidth constraints can be further explored.
    \item[(2)] This paper does not delve deeply into the computational optimization of the proposed method. In fact, for applications in embedded devices such as microcontrollers, radar processors, etc., computational acceleration of the algorithm is crucial. Further efforts should focus on optimizing the computation and storage of the GRAM matrix, as well as numerical algorithms like eigenvalue decomposition.
    \item[(3)] Future work will provide detailed parameter tuning guidelines and attempt to integrate machine learning techniques for adaptive mode decomposition.
\end{itemize}

\bibliographystyle{unsrt} 
\bibliography{sample} 
\end{document}